\documentclass[10pt,journal,compsoc]{IEEEtran}
\usepackage{mathptmx} 
\usepackage{fancyhdr}
\usepackage[normalem]{ulem}
\usepackage[hyphens]{url}
\usepackage[final]{microtype}
\usepackage[keeplastbox]{flushend}
\usepackage[bookmarks=true,breaklinks=true,letterpaper=true,colorlinks,linkcolor=black,citecolor=blue,urlcolor=black]{hyperref}
\usepackage[english]{babel}
\usepackage[utf8]{inputenc}
\usepackage{cite}

\usepackage{amsmath,amssymb,amsfonts}
\usepackage{graphicx}
\DeclareGraphicsExtensions{.pdf,.png,.jpg}
\usepackage{textcomp}
\usepackage{xcolor}
\usepackage{booktabs}
\usepackage{array}
\usepackage{caption}
\usepackage{subcaption}
\usepackage{setspace}
\usepackage{algorithm,algpseudocode}
\algtext*{EndWhile}
\algtext*{EndIf}
\algtext*{EndFor}
\usepackage{balance}
\usepackage{enumitem}
\usepackage{comment}
\usepackage{xcolor}
\begin{document}

\title{Accelerating Large-Scale Graph-based Nearest Neighbor Search on a Computational Storage Platform}

\author{Ji-Hoon Kim,~\IEEEmembership{Graduate Student Member,~IEEE,}
        Yeo-Reum Park,~\IEEEmembership{Graduate Student Member,~IEEE,}
        Jaeyoung Do,~\IEEEmembership{Member,~IEEE,}
        Soo-Young Ji,
        and Joo-Young Kim,~\IEEEmembership{Senior Member,~IEEE}
\IEEEcompsocitemizethanks{\IEEEcompsocthanksitem Ji-Hoon Kim, Yeo-Reum Park, and Joo-Young Kim are with the Department of Electrical Engineering, Korea Advanced Institute of Science and Technology, Daejeon 34141, Republic of Korea. E-mail: \{jihoon0708, summerpark, jooyoung1203\}@kaist.ac.kr
\IEEEcompsocthanksitem Jaeyoung Do is with Amazon Alexa AI, Seattle, WA, USA. This work was done while the author was a member of the database research group at Microsoft Research. E-mail: domjae@amazon.com
\IEEEcompsocthanksitem Soo-Young Ji is with the Advanced Solution Team, Memory Business, Samsung Electronics, Gyonggi-do 18448, Republic of Korea. E-mail: s.young.ji@samsung.com}
\thanks{Manuscript received July 19, 2021; revised January 7, 2022; accepted February 27. This work was supported by the MSIT(Ministry of Science and ICT), Korea, under the ITRC(Information Technology Research Center) support program(IITP-2020-0-01847) supervised by the IITP(Institute for Information \& Communications Technology Planning \& Evaluation), and Samsung Electronics Co., Ltd(IO201210-07991-01). (Corresponding authors: Ji-Hoon Kim; Joo-Young Kim)}}

\markboth{IEEE TRANSACTIONS ON COMPUTERS,~Vol.~XX, No.~XX, Month~Year}%
{Shell \MakeLowercase{\textit{et al.}}: Bare Demo of IEEEtran.cls for Computer Society Journals}

\IEEEtitleabstractindextext{%
\begin{abstract}
$K$-nearest neighbor search is one of the fundamental tasks in various applications and the hierarchical navigable small world (HNSW) has recently drawn attention in large-scale cloud services, as it easily scales up the database while offering fast search. On the other hand, a computational storage device (CSD) that combines programmable logic and storage modules on a single board becomes popular to address the data bandwidth bottleneck of modern computing systems.
In this paper, we propose a computational storage platform that can accelerate a large-scale graph-based nearest neighbor search algorithm based on SmartSSD CSD. To this end, we modify the algorithm more amenable on the hardware and implement two types of accelerators using HLS- and RTL-based methodology with various optimization methods. In addition, we scale up the proposed platform to have 4 SmartSSDs and apply graph parallelism to boost the system performance further. As a result, the proposed computational storage platform achieves 75.59 query per second throughput for the SIFT1B dataset at 258.66W power dissipation, which is 12.83x and 17.91x faster and 10.43x and 24.33x more energy efficient than the conventional CPU-based and GPU-based server platform, respectively. With multi-terabyte storage and custom acceleration capability, we believe that the proposed computational storage platform is a promising solution for cost-sensitive cloud datacenters.
\end{abstract}

\begin{IEEEkeywords}
ANN search, Architecture, Cloud, Data-center, FPGA, In-storage computing, Near-memory computing, SmartSSD.
\end{IEEEkeywords}}

\maketitle
\IEEEdisplaynontitleabstractindextext
\IEEEpeerreviewmaketitle

\IEEEraisesectionheading{\section{Introduction}\label{sec:introduction}}
\IEEEPARstart{T}{he} nearest neighbor search that finds a certain number of closest points in a high dimensional space for a given query is a fundamental task in many domains, including machine learning \cite{wiebe2018quantum}, data analysis \cite{tversky1986nearest}, and information retrieval \cite{duda2001pattern}. As the volume of data available increases exponentially in the era of big data \cite{cai2015challenges}, the importance of a scalable and efficient data search becomes even more significant. For this reason, instead of performing an exhaustive brute-force search to find exact nearest neighbors, approximate nearest neighbors (ANN) search that finds highly probable nearest neighbors with less computational complexity and smaller memory footprint has been widely used in many applications such as image classification \cite{boiman2008defense}, image recognition \cite{torralba2008small}, image search \cite{jegou2010improving}, and text retrieval \cite{xiong2020approximate}.

A common mechanism for the search starts with transforming a large dataset into high-dimensional, real-valued feature vectors by employing a set of machine learning techniques and storing them in a database \cite{huang2013learning}. Because such feature extraction techniques produce similar vectors for similar data, distances (e.g., Euclidean) among the vectors can be measured to quantify the semantic similarity of the corresponding data \cite{chomboon2015empirical}. When a user query is given, the query is first transformed to a feature vector (by using the same feature extraction technique), and then an ANN search is performed to find a list of ranked database vectors that are semantically close to the query vector.

Major search engines, including Bing and Google, also employ a family of ANN search algorithms to boost their search performance \cite{MS}\cite{41694}. However, such cloud services with a large-scale search often pose strong service-level agreement (SLA) requirements such as low latency, high bandwidth, and search relevance. To meet these requirements, cloud service providers need to employ an expensive compute cluster consisting of high-end CPUs with many DIMMs to store the entire database vectors in memory and perform the fast ANN search in a distributed manner. However, this scale-out approach is not only costly from the perspective of operating cost but also expensive due to the time and energy overhead required for moving a huge volume of data from storage to compute devices.
Even in a single node, accelerating ANN search is challenging as the algorithms demand both high memory capacity and bandwidth. Over a few TB storage is needed to store large-sized vector datasets, but the SSD bandwidth is significantly insufficient to match the computing capacity. We observe that the time spent for IO accounts for more than 70\% of the total latency when we run a state-of-the-art proximity graph-based ANN algorithm in a standard CPU-based server system. This bottleneck indicates that the current computer system architecture is not suitable for accelerating large-scale ANN search.

In this paper, we propose a cost-effective and energy-efficient near-data processing (NDP) solution for accelerating a large-scale ANN search for cloud services. In particular, we use a U.2 form-factor flash solid-state drive with Xilinx UltraScale+ FPGA integrated. By placing computing capabilities directly on the storage, data is allowed to be processed in place with significantly less data movement. In addition, this approach frees up host resources such as CPU, DIMM, and network bandwidth, which can be potentially used for other activities running on the host. The main contributions of the paper are as follows.

\begin{itemize}[leftmargin=*]
\item {We implement the state-of-the-art proximity graph-based nearest neighbor search on the computational storage platform for the first time for cloud services.}
\item {We propose a software/hardware co-design approach to make the target graph-based algorithm more suitable for the computational storage platform.}
\item {We accelerate the graph-traversing search kernel that requires a lot of random memory accesses using various hardware optimization techniques. We propose two types of accelerators based on high-level synthesis (HLS) and RTL.}
\item {We scale up the proposed platform to have multiple CSDs that run in parallel and show a linear performance improvement as the number of devices increase.}
\item {We evaluate the overall system performance of the proposed platform for a large-scale dataset and compare it against a conventional CPU-based and GPU-based server platform.}
\end{itemize}
\section{Algorithmic Background}
We first define the nearest neighbor search problem and introduce state-of-the-art ANN search algorithms. We then describe the latest graph-based algorithm named hierarchical navigable small world (HNSW), which is our target algorithm to accelerate on the computational storage platform.

\subsection{Nearest Neighbor Search Definition}
Given a dataset $X$ that consists of $n$ points in $d$ dimensional space, i.e., $X = \left \{ x_{1}, x_{2}, \cdots, x_{n} \right \}$, the nearest neighbor search problem is to find $K$ closest points in $X$ to $q$ based on similarity measure $p(x_{i},q)$, in which $x_{i}$ is a $d$ dimensional vector and a query point $q$. The brute-force search method computes the distance between the query and every point of the dataset guarantees the exact solution, but it is not feasible for a large $n$ value as the computational time increases linearly, $O(n)$. To address this issue, approximate nearest neighbor algorithms that can practically accelerate the search process by two-to-three orders of magnitude are widely used. Their goal is to reduce the computational complexity and the memory footprint while achieving a high probability of exact solution measured in recall metric (= the number of correct neighbors / the total number of neighbors searched).

\subsection{Tree-based Space Partitioning}
Tree-based ANN search algorithms such as KD-trees \cite{friedman1977algorithm}, R-trees \cite{guttman1984r}, and VP-trees \cite{yianilos1993data} have been proposed to accelerate the search process by partitioning the space hierarchically into a tree structure and splitting the points along with the structure. Although these methods reduce the search time substantially for low dimensional datasets, their performance starts to suffer as the number of dimensions increases because distances among points are not distinctive enough \cite{beyer1999nearest}.

\begin{figure}[t] 
    \centering 
    \includegraphics[scale=0.7]{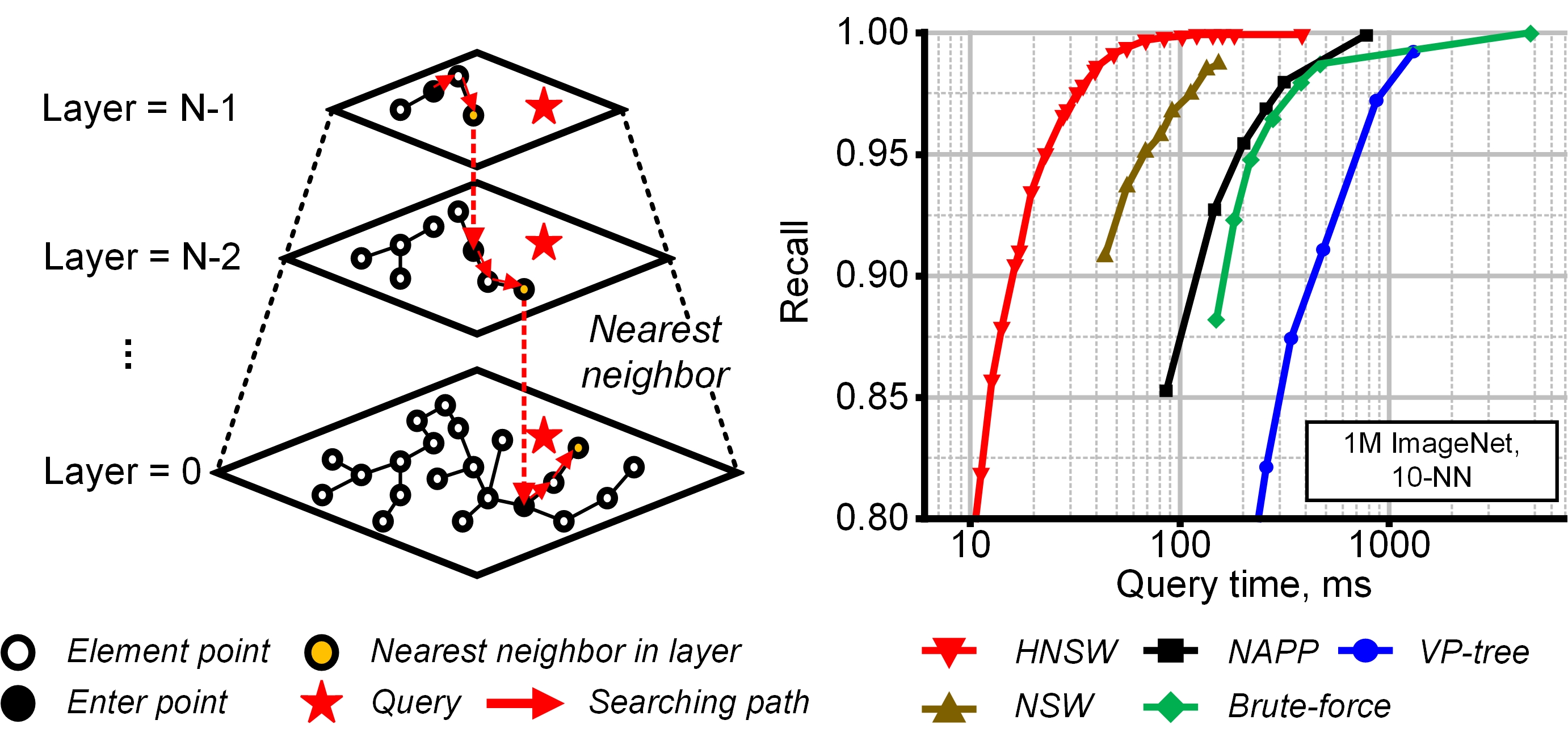} 
    \caption{HNSW's Graph Structure and Search Performance (HNSW\cite{malkov2018efficient}, NAPP\cite{tellez2011succinct}, VP-tree\cite{yianilos1993data}, NSW\cite{dong2011efficient})}
    \label{fig_1} 
\end{figure}

\subsection{Hashing-based Methods}
Hashing-based ANN search algorithms are derived from the idea of locality-sensitive hashing (LSH) \cite{gionis1999similarity} that hashes similar inputs into the same buckets with high probability. This method usually predefines multiple hash tables to convert a vector to shorter hash codes and matches the query against candidate vectors in the hash code domain for efficient computation. In order to improve search accuracy, a lot of literature has been published to design better hash functions, including the learning-based approach \cite{liu2012supervised} and project-based approach \cite{sun2014srs}.

\subsection{Quantization-based Methods}
Product quantization (PQ) based methods \cite{jegou2010product} have been extensively researched in recent years as they reduce the memory footprint significantly by compressing the candidate vectors via a vector or sub-vector quantization. The distance calculation between a query and candidates is also approximated by the distance calculation between the query and the representative words based on the quantization result. Hardware implementations using this method exist due to its low computation and memory requirements \cite{zhang2018efficient}, but they do not output high search quality because of the accuracy loss caused by vector quantization. 

\begin{figure*}[t] 
    \centering 
    \includegraphics[scale=0.7]{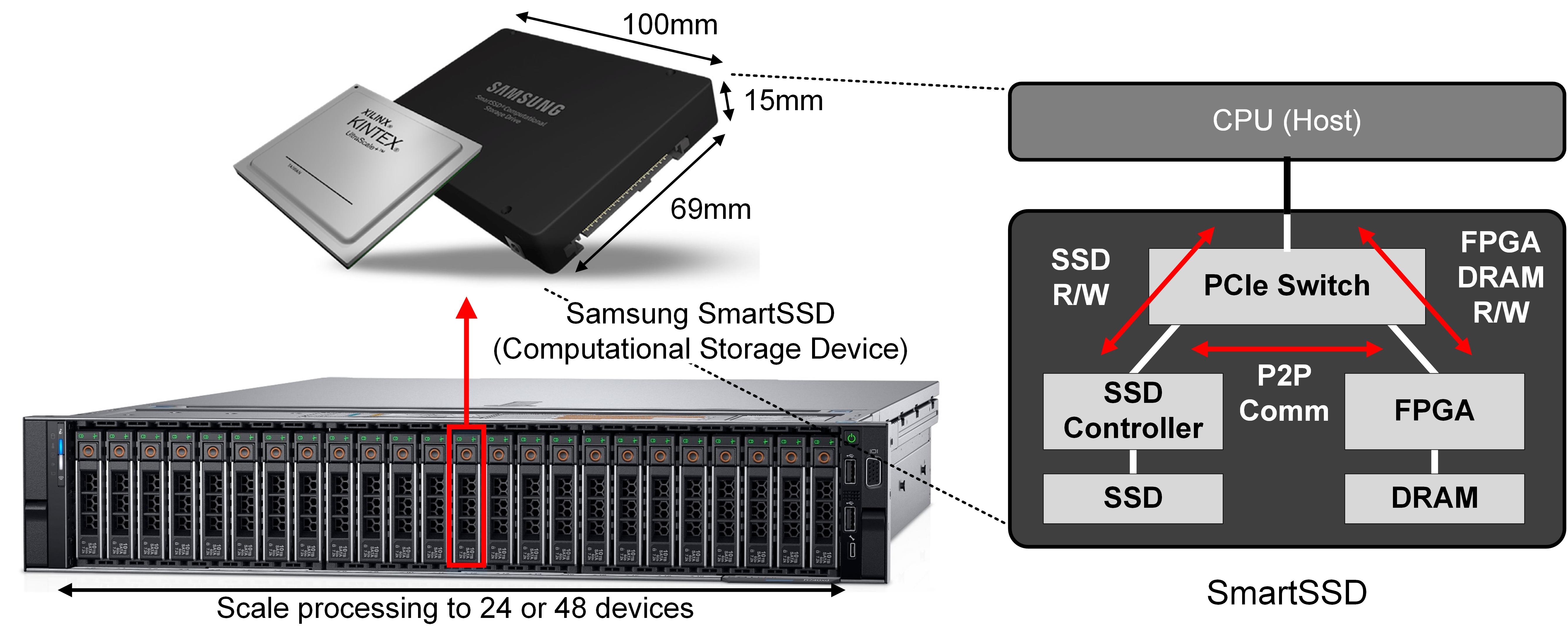} 
    \vspace{-0.1in}
    \caption{SmartSSD System Architecture} 
    \label{fig_2} 
\end{figure*}

\subsection{Graph-based Methods}
Another popular ANN search algorithm is the graph-based approach \cite{dong2011efficient}\cite{malkov2014approximate}. These algorithms build a proximity graph where a candidate vector is represented as a vertex, and two vertices are connected if they are sufficiently close. Then, they search nearest neighbors from an entry point by exploring the graph based on the distance relation among neighbor nodes based on the idea that a neighbor’s neighbor is also likely to be a neighbor. The proximity graph-based methods achieve both high accuracy and fast search, but they have memory space overhead to store original vectors and an additional graph structure. They also require many random accesses for graph traversing, which is unsuitable for memory devices such as DRAM and SSD.

\begin{algorithm}[t]
\caption{HNSW Search Kernel}
\label{algorithm}
\begin{algorithmic}
\footnotesize
\setstretch{0.93} 
\item
{\bf{Input:}} query $q$, enterpoint $ep$, \# of nearest to return $ef$, layer number $l$
\\
{\bf{Output:}} $ef$ closest neighbors to $q$
\\
{\bf{Parameter:}} \textit{maxM}: max \# of lists in upper layers, \textit{maxM0}: max \# of lists in layer0, \textit{V}: visited list, \textit{C}: candidate list, \textit{F}: final list
\begin{algorithmic}[1]
\State{$V, C, F\leftarrow ep$}
\While{$\left | C \right |> 0$}
    \State{$c\leftarrow$ get nearest element from $C$ to $q$}
    \State{$f\leftarrow$ get furthest element from $F$ to $q$}
    \If{distance$\left ( c,q \right )$ $>$ distance$\left ( f,q \right )$}
        \State{\bf{break}}
    \EndIf
    \For{$e\in neighborhood(c)$ at layer $l$}
        \If{$e\notin V$}
            \State{$V\leftarrow V\cup e$}
            \State{$f\leftarrow$ get furthest element from $F$ to $q$}
            \If{distance$\left(e,q \right)$ $<$ distance$\left(f,q \right)$ or $\left | F \right |<ef$}
                \State{$C\leftarrow C\cup e$}
                \State{$F\leftarrow F\cup e$}
                \If{$\left | F \right |>ef$}
                    \State{remove furthest element from $F$ to $q$}
                \EndIf
            \EndIf
        \EndIf
    \EndFor
\EndWhile
\State{\bf{return} $F$}
\end{algorithmic}
\end{algorithmic}
\end{algorithm}

\subsection{Hierarchical Navigable Small World}
Among proximity graph-based algorithms, hierarchical navigable small world (HNSW) \cite{malkov2018efficient} is a target algorithm of this work. The HNSW algorithm has recently gained a lot of attention because of its outstanding performance on high dimensional datasets when compared to previous algorithms. As shown in Figure \ref{fig_1}, the key idea of HNSW algorithm is to separate the links according to their length scale into different layers and then search in a multi-layer graph. It maintains a few long-range connections on the top layer while high-volume and short-range connections lie on the lower layers. In the HNSW algorithm, the search always starts from the coarse-grained top layer which has the longest links. The algorithm greedily traverses through the elements from the upper layer until a local minimum is reached. After that, the search switches to the lower layer which has shorter links and restarts the process from the element which was the local minimum in the previous layer. This process is repeated until reaching to the finest-grained bottom layer. As a result, the HNSW's unique graph structure and search method enable much better logarithmic complexity scaling and faster search speed than the other previous algorithms (Figure \ref{fig_1}). 

Since the nearest neighbor search for user queries are conducted through the stored HNSW graph database which was already constructed in a downtime, the focus of this work is the search kernel of the HNSW algorithm. The Algorithm \ref{algorithm} describes the HNSW search kernel and its essential parameters. When the user query $q$ is given, it traverses the graph from the top layer starting at the entering point $ep$ and tries to find the $ef$ closest neighbors in the vector space.
The $ef$ is the main parameter that controls the quality of the search. It determines the size of the buffer window that tracks the closest neighbors, while it also affects how many candidates the algorithm needs to go through. The final $K$ nearest neighbors are simply selected among the $ef$ closest neighbors.
The search kernel chooses the entering point of the next layer based on the closest neighbors found in the current layer and continues the search on the next layer. The kernel repeats this process until it reaches the bottom layer that contains all the points in the dataset.

In order to find the closest neighbors, the kernel utilizes three lists: visited list $v$, candidate list $c$, and final list $f$. The visited list, whose size is the same as the number of total points, is used to check if a point is visited before, so the same distance calculation is not repeated. The candidate list is the list of candidate points for both graph traversing and closest neighbors. Its size is set to a larger number than $ef$. The list also tracks the closest point to the query among the candidate points along with its distance value. The final list is a dynamic list that only keeps the $ef$ closest neighbors updated. Unlike the candidate list, it tracks the furthest point in the list. 
All three lists are initialized to have only the entering point. The search kernel reads a candidate from the candidate list and visits its neighbor points one by one. If the distance between the query and the visited neighbor is smaller than the maximum distance of the final list, the neighbor point is inserted into both the candidate and final list. If the final list is full, it removes the current furthest and inserts the new point with the maximum distance updated. The kernel repeats this process until there is no candidate left or until the minimum distance of the candidate list is no longer smaller than the maximum distance of the final list. In other words, the search kernel stops traversing if no closer point can be found in the neighbor points of a candidate.
\section{Computational Storage Device}
As the volume of machine and user-generated data exponentially increases in the big data era, the demand for high-capacity storage has been increasing, especially in cloud data centers. At the same time, the high volume of data has created challenges for reliable storage and efficient data retrieval for further processing. Current server architectures, broadly composed of CPU and storage, move all necessary data from the storage to the host memory when they need processing. As both components constantly improve their performances, bulk data transfer between the CPU and storage causes significant delays in the system. To solve this memory bandwidth bottleneck issue, Samsung developed SmartSSD \cite{smartssd} computational storage device that allows parallel computations on the storage device without data being moved to the host memory. SmartSSD combines a field-programmable gate array (FPGA) and a solid-state drive (SSD) with a fast direct data movement channel between the two for efficient near-data processing. Evolved from the first HHHL PCIe form factor, Samsung recently released a more compact version with the 2.5 inches U.2 form factor that is interchangeable with other U.2 SSD cards. The major advantage of using computational storage is offloading computations to the storage level and reducing the volume of data fetched to the CPU due to its near-data processing capability. In other words, computational storage reduces the memory bandwidth requirement between the CPU and storage, which is often the source of system bottleneck, by allowing the CPU to read-only extracted information out of bulk storage data. Many previous works \cite{do2013query}\cite{kang2013enabling}\cite{gu2016biscuit}\cite{lee2020smartssd} discussed the advantages of computational storage.

 \begin{figure}[t] 
    \centering 
    \includegraphics[scale=0.7]{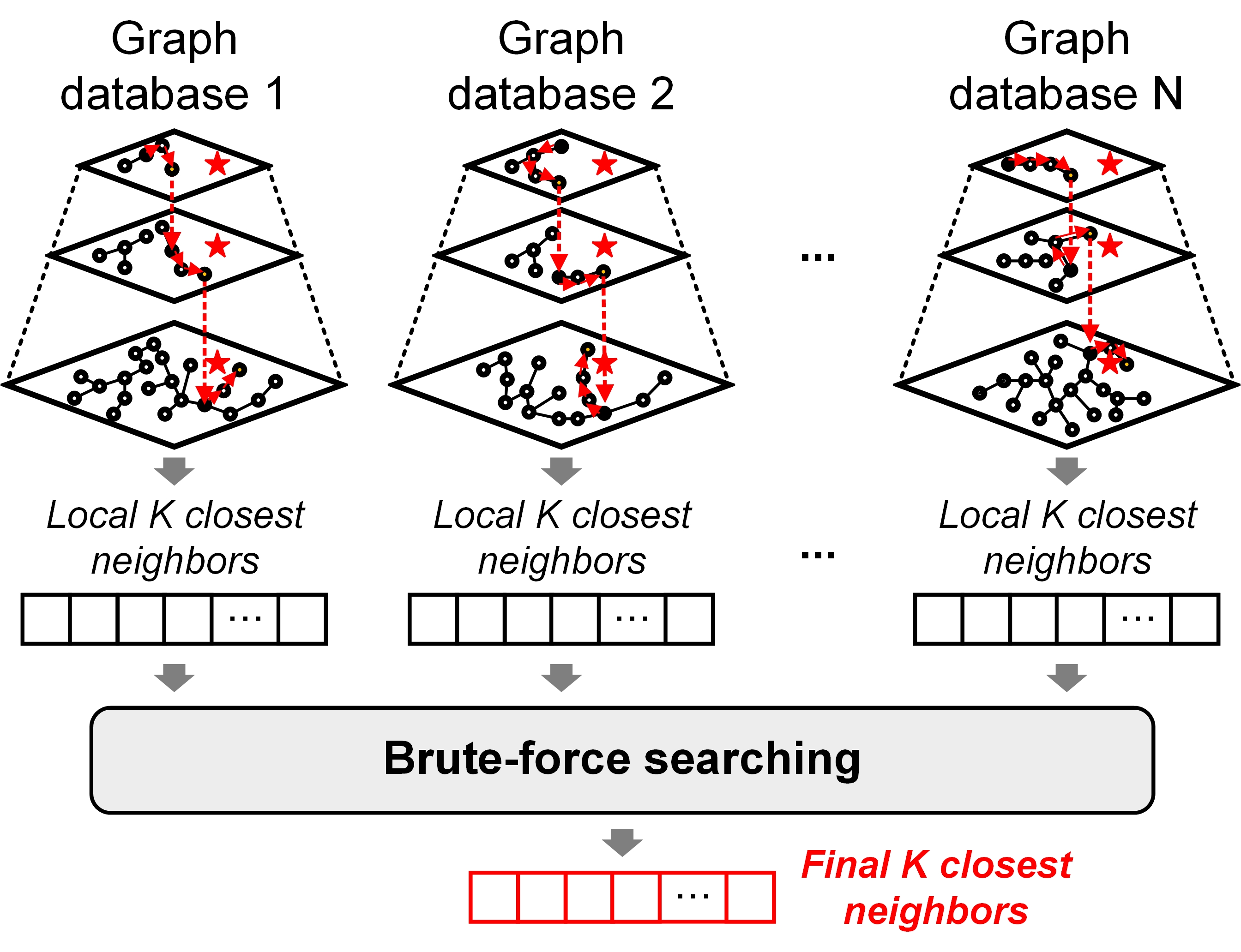}
    \caption{HNSW Algorithm Modification} 
    \label{fig_3}
\end{figure}

Figure \ref{fig_2} shows the system architecture of the SmartSSD platform. It has a Xilinx Kintex UltraScale+ KU15P FPGA with 4GB DRAM and 3.84TB NAND Flash arrays on the same board in U.2 form factor ($69mm\times100mm\times 15mm$).
The FPGA chip contains 1.14 million logic cells, 1968 DSP slices, and 34.6 Mbits on-chip SRAM. The SmartSSD device is connected to the host CPU through the PCIe Gen3x4 interface, which gives a theoretical maximum bandwidth of 4 GB/s. One of the main features of the SmartSSD platform is that there is a PCIe switch presented in the FPGA to provide three-way paths: between the CPU and FPGA, between the CPU and SSD, and between the FPGA and SSD. The 4GB DRAM attached to the FPGA can work as a buffer memory when data moves between FPGA and CPU or between FPGA and SSD because the FPGA does not have enough memory on the chip. 
From the host driver’s view, each FPGA and SSD is seen as a PCIe device. The host transfers data between the host memory and FPGA through normal PCIe communication. It also transfers data between the host memory and SSD in the same way, even though the data physically go through the PCIe switch on the FPGA. SmartSSD also supports direct data transfer between the FPGA and SSD, called peer-to-peer (P2P) communication, because it does not involve any memory copy on the host side. For P2P communication, the driver uses a reserved memory space on the FPGA’s DRAM called a common memory area (CMA). This special memory area is made to be accessible by the host, FPGA, and SSD, and it is exposed to the host PCIe address map. The host application cannot directly access the CMA but needs to go through Xilinx’s OpenCL APIs \cite{fifield2016optimizing}. The application allocates memory from CMA using \textit{clCreateBuffer} API with Xilinx memory extension and P2P flag set. The memory object returned from the previous operation is passed to \textit{clEnqueueMapBuffer} API, which provides a virtual address pointer. Then, the SSD can use the allocated memory in the CMA for direct read/write access through the obtained virtual address. This process creates the P2P communication between the FPGA and SSD and removes unnecessary host interaction and memory copies.

Since the SmartSSD was developed in U.2 form factor, it is easy to scale up to have multiple cards using a commercially available storage server. For the case of 48 SmartSSD cards, the server's total storage capacity will be 184.32TB with more than 50 million logic cells and 192GB DRAMs. In this work, we harness 4 SmartSSD devices in a single storage server and prove the scalability of the system with a non-trivial graph-based nearest neighbor search application. Furthermore, we suggest that this computational storage platform is a high-performance and cost-effective solution for datacenters because it uses near-data processing and reduces the number of nodes with high storage capacity.
\section{Software-Hardware Co-Design}

\begin{figure}[t] 
    \centering 
    \includegraphics[scale=0.7]{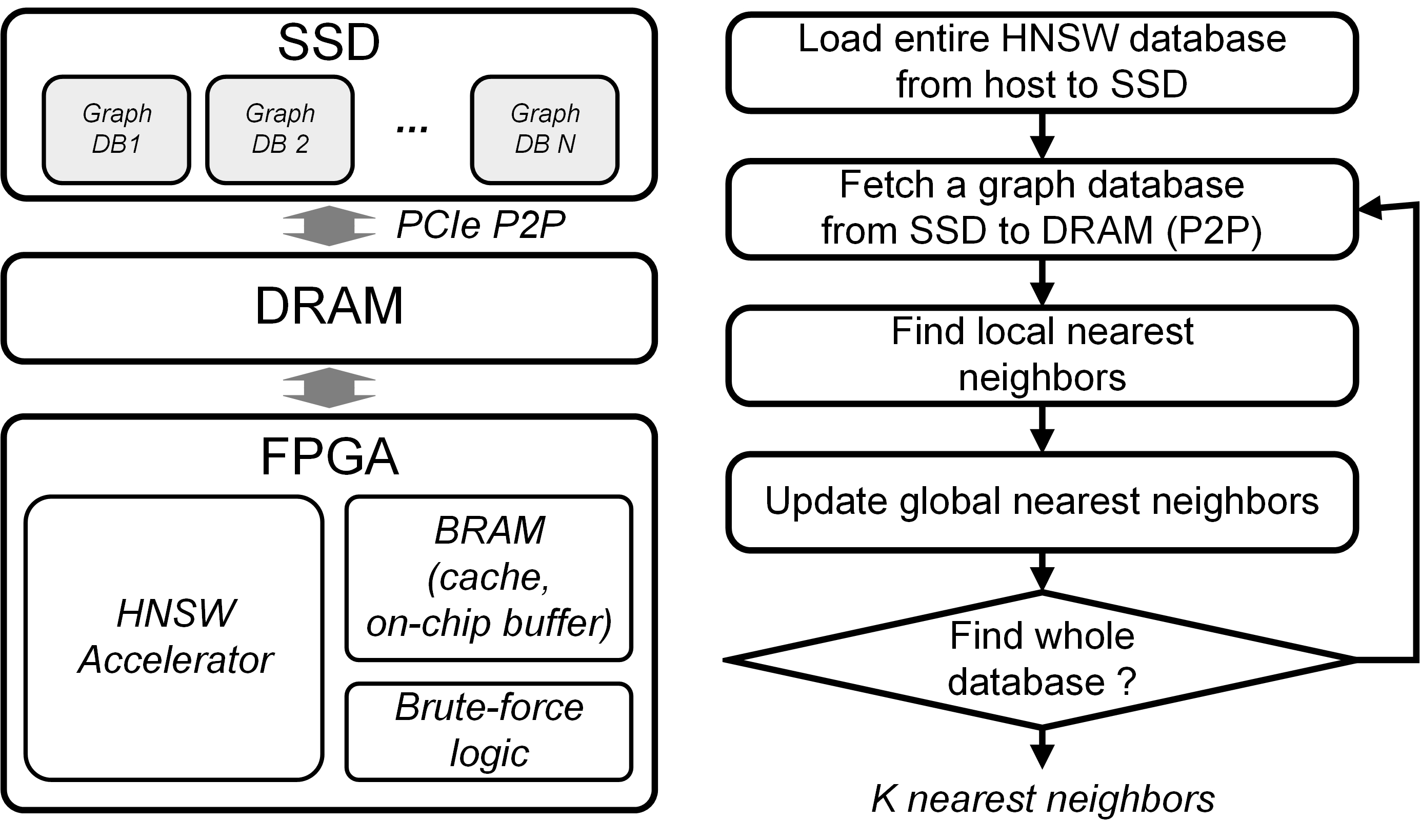}
    \caption{Hardware Mapping and Dataflow} 
    \label{fig_4} 
\end{figure}

\subsection{Algorithm Modification}
As stated in section 2.6, the HNSW algorithm builds a multi-layer proximity graph that requires lots of random data accesses for traversing points in the graph. We can utilize DRAM’s fast accesses for small datasets whose graph database can fit the FPGA’s 4GB DRAM. However, for large datasets whose sizes easily extend to a few hundred gigabytes (e.g., SIFT1B), DRAM is not feasible. Leveraging SmartSSD’s large storage capacity, we can store the database to the SSD and perform the search on the FPGA. The one disadvantage of using an SSD is its long access time.
To mitigate this issue, we modify the HNSW algorithm into two stages, as shown in Figure \ref{fig_3}. Instead of having a monolithic graph database whose size can increase far beyond the FPGA’s DRAM size, we partition it into multiple databases so that each of them is a smaller HNSW graph database that can fit in the DRAM. In detail, we split the raw dataset into $N$ segments first and generate the HNSW database from each of them to make sure its size is less than the size of DRAM. For nearest neighbor search, we run an independent search on each graph database for a given query and get $N$ sets of $K$ closest neighbors as results. Then, we perform brute-force distance calculations for all intermediate results to find the final $K$ closest neighbors. This two-stage algorithm modification shows good performance and does not suffer accuracy loss because each HNSW graph is large enough to produce relevant results in the first stage, and those results are reduced with an exhaustive method in the second stage. For the SIFT1B dataset, the recall of the modified HNSW is 0.94 when \textit{K=10} with \textit{ef=40}.

\begin{figure*}[t] 
    \centering 
    \includegraphics[scale=0.6]{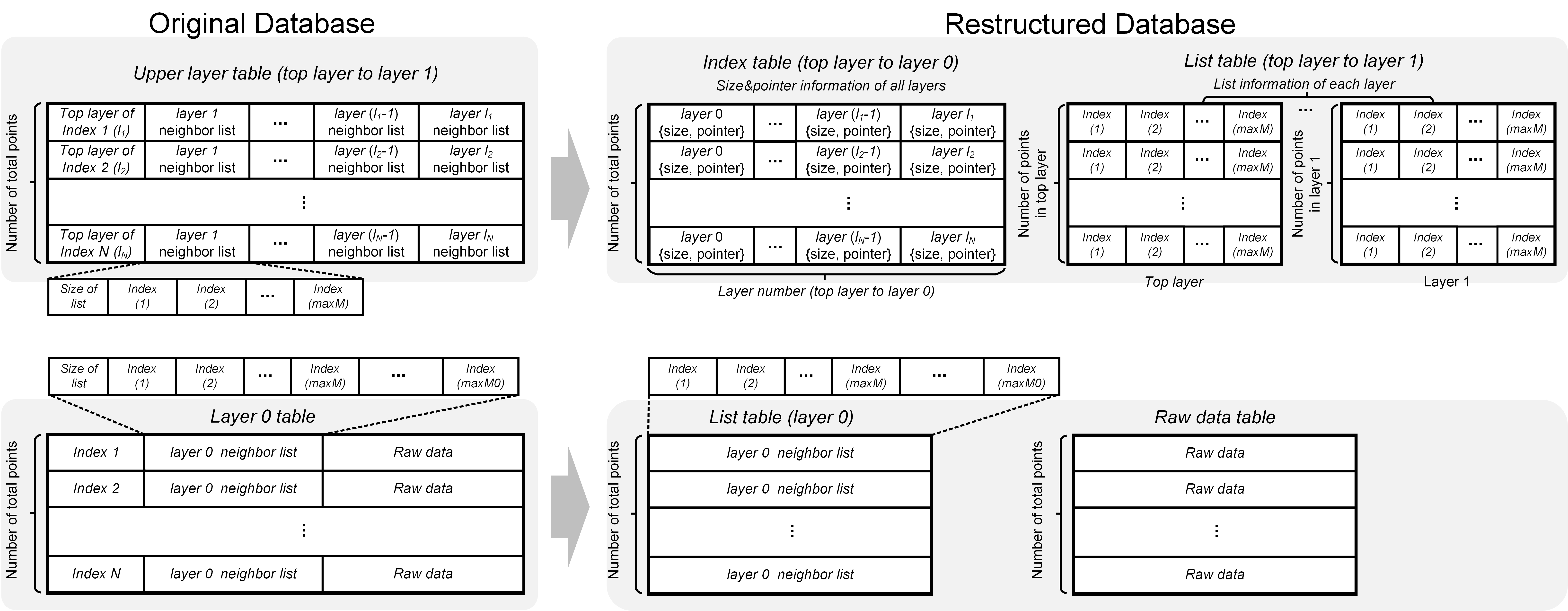} 
    \caption{Database Restructuring} 
    \vspace{-0.1in}
    \label{fig_5} 
\end{figure*}

\subsection{Hardware Mapping}
Due to the slight modification of the algorithm, we can efficiently map the HNSW algorithm on the SmartSSD platform using all the memory components, such as on-chip BRAM, off-chip DRAM, and SSD. Figure \ref{fig_4} describes the overall hardware mapping and the data flow. First, the host loads the entire database that includes multiple HNSW graph databases to the SSD. This is a single-time event that initializes the SSD and is the baseline of our platform. Second, the FPGA fetches a graph database to the DRAM using P2P communication and searches for the input queries. It finds the $K$ nearest neighbors for the current graph database and updates the final nearest neighbors that keep the best results among all the nearest neighbors found so far. Third, the FPGA repeats this procedure until it covers all the graph databases. Once it finds the final nearest neighbor results, it sends them back to the host. By doing this, the FPGA fetches the data from the SSD to DRAM only once for each graph database. It also uses P2P communication, which eliminates unnecessary PCIe traffic to the host and helps to improve system performance.

\subsection{Database Restructuring and Caching}
The original database of the HNSW algorithm consists of two tables: 1) an upper-layer table that contains the linkage information of the points in the upper layers and 2) a layer-0 table that contains the linkage information of the points in the bottom layer along with their raw data. Each line of the upper-layer table describes a point’s linkage information with the highest layer that the point is observed and the neighbor lists for all the layers except the bottom. Each layer’s neighbor list starts with the size information that stores the number of linked points to the target point, followed by the indexes of the linked points. The maximum size of each neighbor list is set by the parameter called $maxM$. If a point is only observed in the bottom layer, it stores 0 for the first observed field of the highest layer and omits the neighbor lists in the upper-layer table to save the table size.
On the other hand, the layer-0 table has a fixed size for each point. Each line starts with the index of the point, which is the same as the line number, and includes the size of the neighbor list whose maximum is set to twice $maxM$ $(maxM0 = 2 \times maxM)$. After the neighbor list, the raw vector data follow. 
This original HNSW database structure can store the graph information in a more compact manner, but it requires additional calculations to index the target point's information in upper-layer table and makes unaligned addresses both in upper-layer and layer-0 table. As a result, using the original database structure increases the number of external memory accesses on hardware when graph traversing, which can significantly degrade the overall search performance.
To solve this problem, we re-organize the original database into three types of tables to have a more aligned address for hardware: 1) an index table that stores the size and pointer information to access the following list tables efficiently, 2) the list tables that store the lists of neighbor indexes, and 3) the raw data table that stores the raw vector data, as shown in Figure \ref{fig_5}. Specifically, each line of the index table contains the size of the neighbor list and the pointer to access the neighbor list in the list table for each layer. Hence, if we read a single line from the index table for a point, we can get how to access its neighbor list in each layer along with the list’s size information. Each entry of the list table stores a point’s neighbor list whose maximum size is fixed to $maxM$ in upper layers and $maxM0$ in layer-0. For the layer-0 table, we separate the neighbor list and raw data into different (size information is already in the index table). This separation allows the search kernel to access the list table and raw data more easily, removing all the unaligned accesses in the original structure. Finally, the proposed reorganized database structure can eliminate the redundant external memory accesses by aligning the memory address with 64 bytes, causing only 4\% increase in database size compared to the original database structure.
In addition, we cache some of the restructured tables on-chip for faster database accesses. We initially designed a standard demand-based cache, but the irregular memory access pattern by the algorithm makes it impractical to utilize the temporal and spatial locality. We only got a hit rate of 16.3\% with a 4-way set-associative cache.
Since the number of index table entries is the same as the number of total points, all entries cannot be cached. Instead, we minimize the access to the index table to one time per point by putting all necessary information in a single row. Then, we decide to cache the list tables from the top layer because every search process starts from the top layer. Due to the FPGA’s on-chip memory capacity, we manage to store the list tables from layer 6 (top) to layer 3.
\section{Accelerator Design}
In this section, we describe two types of hardware accelerators that implement the sophisticated HNSW search algorithm based on different design methodologies: high-level synthesis (HLS) based and RTL based.

\subsection{HLS-based Design}
For the baseline implementation of HLS based design, we use HNSW’s original C++ implementation in GitHub \cite{hnswgit}. We modify the code to synthesize on Vivado HLS, such as replacing dynamic memory allocation and data types with static versions. Based on this baseline, the following techniques are applied to maximize performance.

\subsubsection{Single-bit Tag-based Visited List}
The visited list has the same size as the total number of the points as it needs to check if any point of the dataset is visited for each query. In the original C++ code, the data type of the visited list is set to an unsigned integer, and its value is updated to the index of a query if visited. By comparing the current query’s index and the target point’s value in the visited list, the search kernel knows if the point was visited before. This method works well in the software implementation but requires a large memory space when the dataset size is large (e.g., 4MB on-chip memory for a million point dataset). Considering the FPGA’s on-chip memory size is very limited, the large size of the visited list severely hinders hardware scalability. To address this issue, we change the visited list to a single-bit tag-based implementation, where each bit represents whether each point is visited or not. Although this method requires the initialization of the list for each query processing, it reduces the size of required on-chip memory by 32 times.
 
\subsubsection{DRAM Access Optimization}
To maximize the DRAM bandwidth, we change all the data types that access the external DRAM to maximum 512 bits using HLS's arbitrary-precision data type $ap\_uint\left \langle N \right \rangle$. We use the same data type for the on-chip table accesses that also require high data bandwidth. In addition, since the corresponding bit-width is aligned with the address of the reorganized database that we proposed, the redundant external memory access does not happen.

\begin{figure}[t] 
    \centering 
    \includegraphics[scale=0.7]{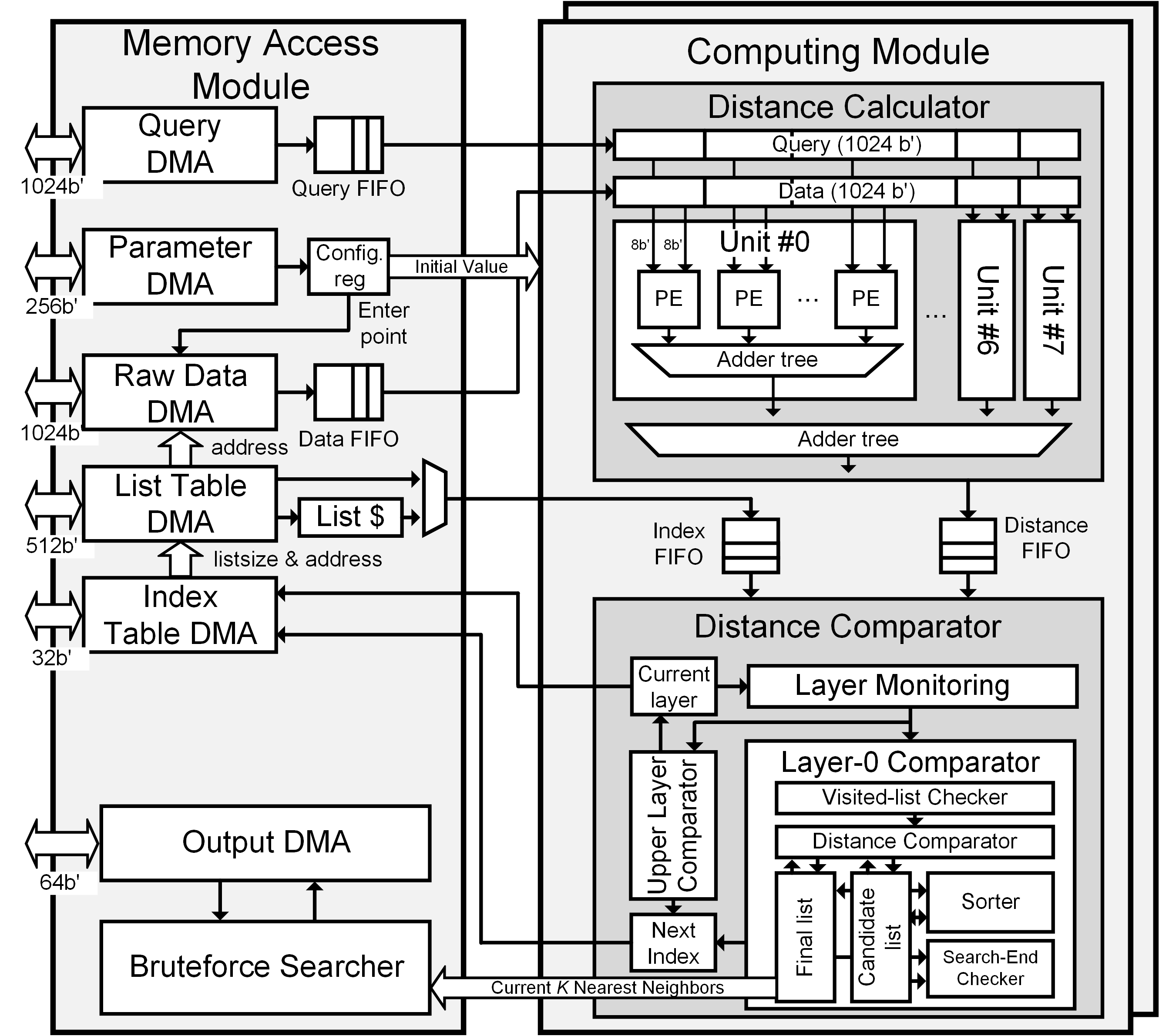} 
    \caption{Overall Architecture} 
    \label{fig_6} 
\end{figure}
 
\subsubsection{Parallelization \& Multi-Query Processing}
On the compute side, the search kernel reads a candidate index from the candidate list and visits its neighbor points stored in the list one by one. During this process, it calculates the distance, updates the three lists, and finds the next candidate. We parallelize this process by reading multiple neighbor points and performing distance calculations in parallel. The search kernel then decides how to update the lists with simple comparisons. We also use the \textit{HLS pipeline} and \textit{HLS unroll} pragma to speed up the distance calculation.
In addition, multiple queries can be processed simultaneously because the search processing for a query is read-only and independent. In HLS, we instantiate multiple search kernel modules to achieve higher throughput. We can successfully integrate two kernel modules with the given FPGA resource.

\subsection{RTL-based Design}
Although the HLS-based design has the advantage of fast development and easy debugging, its performance is often limited. One major problem of HLS-based design is that it is hard to fully utilize the external DRAM bandwidth even after substantial optimizations. Also, the HLS tool often does not support synthesis for custom parallelization schemes over well-known code patterns.
As a result, the HNSW algorithm turns out to be a bad fit for HLS as it requires abundant memory accesses and complex dataflow. Therefore, we also design a custom accelerator in System Verilog to fully exploit the acceleration opportunities in the algorithm.

\subsubsection{Overall Architecture}
Figure \ref{fig_6} shows the overall architecture of the proposed HNSW accelerator. It consists of a memory access module that manages all communications with the external DRAM and computing modules that perform distance calculations and neighbor traversing to search the nearest points in each layer. Since the memory access module and the computing modules are cleanly decoupled via FIFO interfaces, it is easy to increase the number of computing modules to support multi-query processing. We integrate two computing modules like in the HLS version in this accelerator design.

\subsubsection{Upper-layer Operation}
For the operation of the proposed accelerator, the host CPU initializes the SmartSSD's SSD with the whole graph databases.
On the FPGA side, once it gets the start signal from the host, the accelerator loads the parameters (e.g., max layer, entering point, \textit{ef} value, etc.) and input queries from the external DRAM to configuration registers and the query FIFO via the parameter DMA and query DMA, respectively. The raw data DMA then loads the entry point's vector data, and the distance calculator computes the distance between the query and the entering point. 
Once the initial distance between the query and the entering point is calculated, the distance comparator starts the candidate searching. Based on the HNSW algorithm, the candidate searching for upper layers does not use the candidate list, and the other two lists as the \textit{ef} value is set to 1. It is rather done by simple distance comparison. If the calculated distance is smaller than the minimum distance stored in the internal register, the upper layer comparator updates the minimum distance and sends the index value to the index table DMA.
Then, the index table DMA accesses the index table, and the list table DMA fetches the neighbor list of the target point. For all the indices in the neighbor list, the raw data DMA reads the vectors one by one, and the distance calculator calculates distances between the vectors and the query. If the upper layer comparator finds a smaller distance than the minimum value among the calculated distances, the index of the current minimum distance is sent to the index table DMA to traverse the following list in the same layer. If not, the current layer is lowered to the next layer and sends the index of the current minimum distance to the index table DMA to start a new search as the entering point of the new layer. The accelerator repeats this process until it reaches the bottom layer.

\subsubsection{Layer-0 Operation}
For the layer-0 case, the layer-0 comparator performs candidate searching by managing the visited list, candidate list, and final list, as explained in section 2.6. The visited-list checker checks whether the current index has been visited before to prevent duplicated distance calculations. If the index is not visited and its distance is smaller than the maximum distance of the final list, the index and distance value are emplaced in the final and candidate list. 
Both lists are sorted lists, so each requires sorting when the new data comes in. Once the layer-0 comparator finishes all the distance calculations and list updates for the neighbor list of the current index, the search-end condition checker checks if the search should be ended. It stops searching only if the candidate list becomes empty or cannot find any good candidate (i.e., min. distance of candidate list $>$ max. distance of final list). Otherwise, it continues searching in the following index by sending the current minimum index of the candidate list to the index table DMA.

\subsubsection{Memory Access Module}
Each DMA module in the memory access module consists of an address generator that generates a proper address to the target table and an AXI read/write master that access the external DRAM. Each DMA is designed to maximize effective DRAM bandwidth with utmost bit-width based on each table's data type and data length. Especially, the query DMA and raw data DMA are designed with the AXI's maximum 1024 bits to support bulk data read of high dimensional vectors. Since all DMA modules always access the restructured database with aligned address, the memory access module can minimize the number of external memory accesses.

\subsubsection{Distance Calculator}
The distance calculator calculates the Euclidean distance between a given query and a vector. A single distance calculation unit comprised of 16 processing elements (PEs) and an adder tree. Each PE calculates the square of the difference between the two 8-bit elements and the adder tree accumulates all the square products in the unit. As the distance calculator integrates 8 calculation units and another adder tree at the bottom, it can compute the distance between the two 128-dimensional vectors in parallel at a time.

\subsubsection{Distance Comparator}
\begin{figure}[t] 
    \centering 
    \includegraphics[scale=0.55]{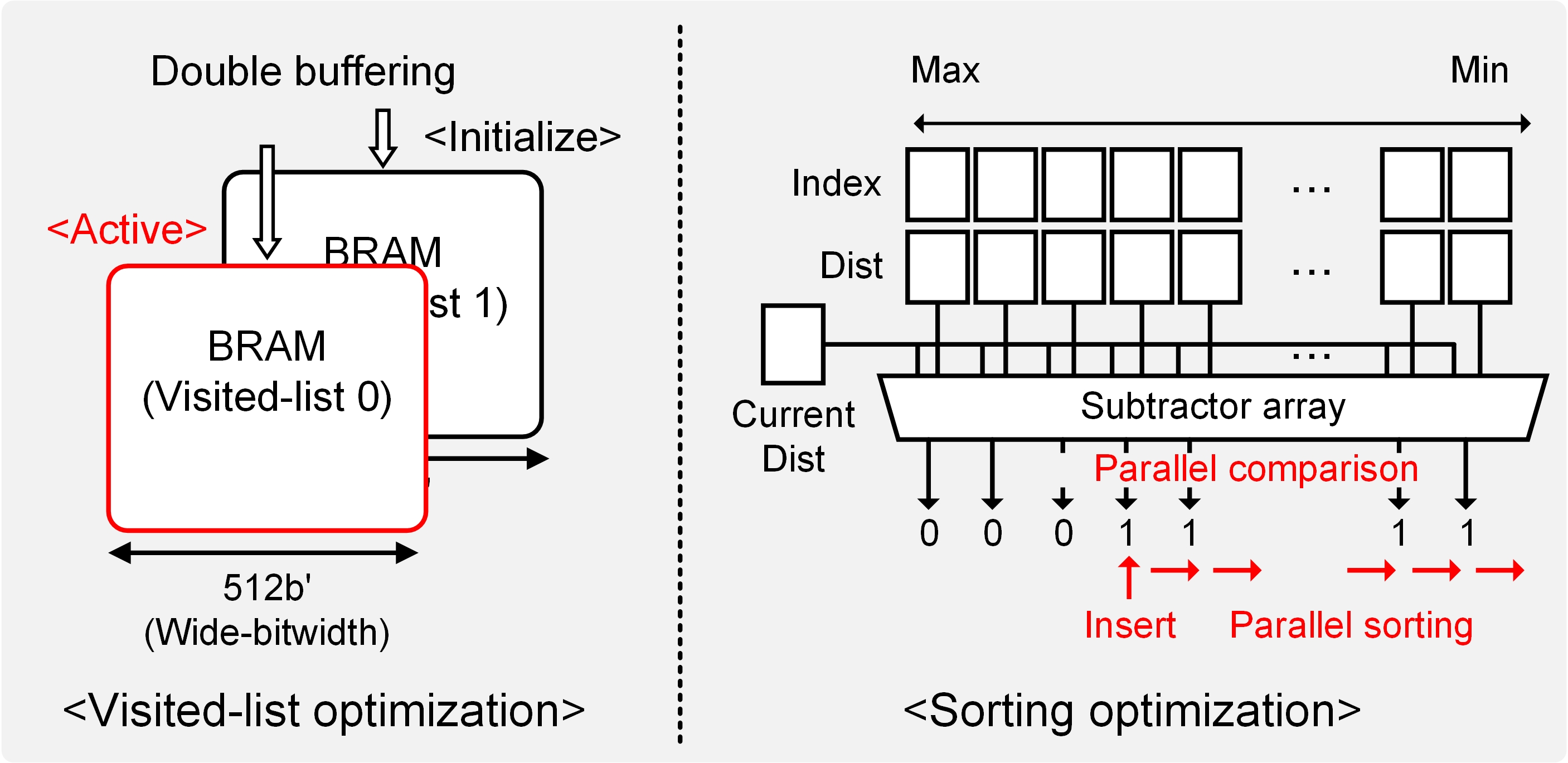} 
    \caption{RTL-based Design Optimizations } 
    \label{fig_7} 
\end{figure}

The distance comparator is responsible for searching candidate indexes and keeping the $ef$ nearest neighbors to return the final $K$ nearest neighbors. In this iterative process, initialization of the single-bit tag-based visited list and sorting of the candidate and the final list is crucial to the performance because the former happens for each new query and the latter happens for each new index insertion, respectively. To address this, we propose two design optimizations illustrated in Figure \ref{fig_7}.
In our implementation, the byte size of the visited list is set to 0.62MB to support a total of 5M points. To minimize the number of cycles for re-writing all the contents to 0, we shape on-chip BRAMs as wide as possible to the maximum of 512 bits. We also employ a couple of visited lists to hide the latency by using one as the active list while the other is initialized. 
In order to speed up the list sorting, we implement the parallel sorting instead of the serial sorting used in the original algorithm. We subtract the current distance from each distance of the list in parallel. Having a bit vector as a result, in which each bit indicates whether the current distance is smaller than the corresponding index (=0) or not (=1), we can determine where the current distance should be inserted in the list.

\section{Experimental Results}
In this section, we present the implementation results of the proposed HNSW accelerators and the computational storage platform using SmartSSD devices. We then evaluate the proposed platform against other server platforms.

\subsection{Experimental Setup}
We chose three server-class computing platforms for evaluation: CPU-based server platform, GPU-based server platform, and SmartSSD-based computational storage server platform. The CPU-based server has 2 AMD EPYC 7351 CPUs, 128GB DIMM, and 1TB SSD in a 2U server rack. The original C++ HNSW implementation \cite{hnswgit} ran on this server. Since it does not have any additional computing device, the CPU is solely responsible for all the computations and data movements. The GPU-based server has 2 Intel Xeon Gold 6226R CPUs and a TITAN RTX GPU that has 4608 CUDA cores with 24GB GDDR6. We run the latest CUDA accelerated HNSW implementation \cite{cuhnswgit} on the GPU-based server. In this case, the CPU handles data movement while the GPU performs the search computations. The computational storage server has 2 Intel Xeon Silver 4210 CPUs and SmartSSD devices, each containing Xilinx's UltraScale+ FPGA, 4GB DDR4, and 3.84TB SSD as near data processing accelerators. In this platform, CPU's role is minimized to the control of the devices, including device programming and launch of P2P communications and no longer handles large data movement or computation. We vary the number of SmartSSD devices from 1 to 4 and run our HLS-based and RTL-based HNSW accelerator on the computational storage server. The detailed hardware configurations of each server platform are listed in Table \ref{table_1}.
We chose the SIFT1B \cite{sift1b} dataset for benchmark, which is widely used for evaluating large-scale ANN algorithms. The dataset contains 1 billion SIFT vectors and 10K queries. Each vector is 128-dimensional with a byte element format. The total size of the SIFT1B dataset is 119GB.

\subsection{FPGA Implementation Results}
\begin{figure}[t] 
    \centering 
    \includegraphics[scale=0.75]{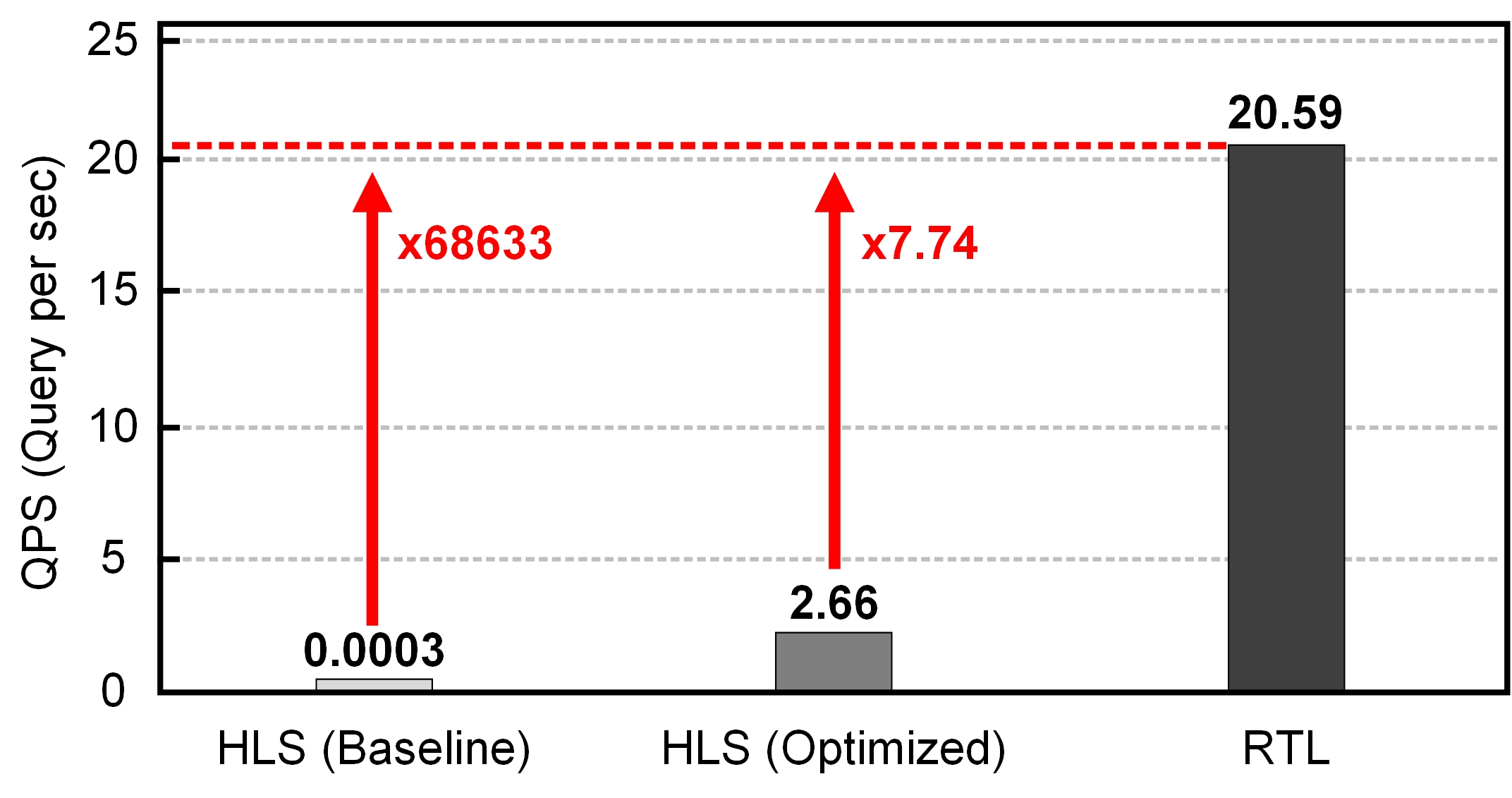}
    \caption{Performance of FPGA Implementations} 
    \vspace{-0.2in}
    \label{fig_8} 
\end{figure}

Table \ref{table_2} shows the FPGA resource utilization of the proposed HLS-based and RTL-based designs. Both designs are bounded by memory resources with more than 80\% usage, mostly spent by the visited list and on-chip list tables. Figure \ref{fig_8} shows the query per second (QPS) performances of the two HLS implementations (baseline and optimized) and RTL implementation. The optimized HLS design achieves 2.66 QPS, which is 8,867 times faster than the baseline HLS design. This performance gain is achieved by database restructuring and bit-width optimization that improved the external memory access pattern. The RTL design achieves 20.59 QPS, improving the optimized HLS design by 7.74 times (68,633x of baseline). This performance gain is achieved by addressing the memory bottleneck still existed in the optimized HLS design. We fully utilize the physical data bandwidth by designing a custom memory interface and minimizing external memory access. This increase of effective bandwidth results in a drastic overall performance improvement.

\begin{figure}[t] 
    \centering 
    \includegraphics[scale=0.7]{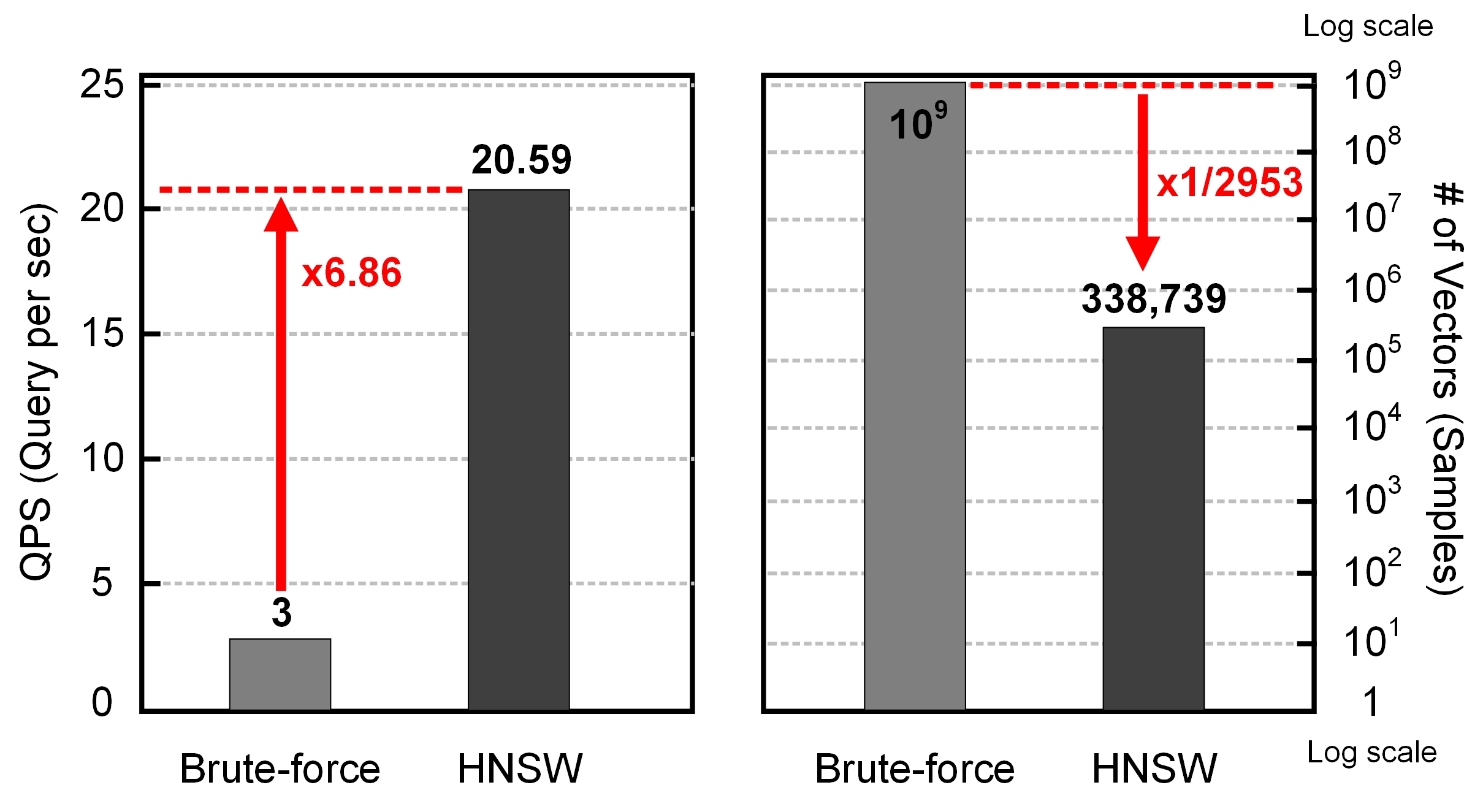} 
    \vspace{-0.1in}
    \caption{Comparison between Brute-force and HNSW Implementation: (a) QPS (b) Number of Vector Reads} 
    \label{fig_9} 
\end{figure}

In addition, we compare the performance of our RTL-based design against the brute-force approach if it was implemented on the SmartSSD's FPGA. The theoretical maximum QPS performance of the brute-force design can be calculated using the number of DSP slices, operating frequency, and the number of vector reads for searching. Since the total number of DSPs in the FPGA is 1968 and each vector is 128-dimensional, the brute-force design would be able to process a maximum of 15 distance calculations between a query and vectors in parallel. Assuming the operating frequency is 200MHz, the computing throughput of the brute-force design is 3 billion vectors per second. A single query needs to compute with 1 billion vectors in the SIFT1B dataset, so it takes 0.33 seconds to process. In this time, 1920 DSPs execute 15 parallel distance calculations every single cycle with 100\% utilization. We do not consider any time for distance comparisons and data fetch from memory. Hence, 3 QPS is the theoretical maximum performance of the brute-force design. Figure \ref{fig_9} shows the QPS and the number of vector visits required for searching in both HNSW (RTL-based) and brute-force design. The HNSW implementation shows a 6.86x higher QPS throughput than the brute-force implementation. The main reason for this performance gap is that the HNSW implementation reduces the number of vectors required in a single query search by 338,739 times, which is 0.03\% of the brute-force case. Although the HNSW requires a lot more complex computation and memory access per vector read, the much smaller number of reads enables the HNSW to outperform the compute-bound brute-force approach.

\begin{table}[t]
\centering
\caption{Experiment Hardware Configurations}
\vspace{-0.05in}
\label{table_1}
{\footnotesize
    \centering
    \resizebox{\columnwidth}{!}{
    \begin{tabular}{cccc}
        \toprule
        {\bf } & {\bf CPU-based} & {\bf  GPU-based} & {\bf SmartSSD-based}\\
        {\bf } & {\bf Server Platform} & {\bf Server Platform} & {\bf Computational Storage Platform}\\
        \hline\hline
                        &  2 AMD        & NVIDIA                   & Xilinx Kintex\\
        Compute Unit    &  EPYC 7351    & TITAN RTX                & UltraScale+KU15P\\
                        &  CPUs         & GPU                      & FPGA\\
        \hline
                        &                                       & GPU Card                              & U.2\\
        Form Factor     & -                                     & {\footnotesize($116mm\times35mm$}     & {\footnotesize($69mm\times15mm$}\\
                        &                                       & {\footnotesize$\times267mm$)}         & {\footnotesize$\times100mm$)}\\
        \hline
        DRAM            & 128GB DIMM       & 24GB GDDR6                  & 4GB DDR4\\
        Capacity        & (DDR4-2666)      &                             & (DDR4-2400)\\
        \hline
        SSD Capacity    & 1TB           & 1TB                   & 3.84TB\\
        \hline 
        SSD             & 0.58GB/s      & 0.58GB/s              & 4GB/s\\
        Bandwidth       & SATA Express  & SATA Express          &NVMe U.2\\
        \bottomrule
    \end{tabular}}
}
\end{table}

\begin{table}[t]
\centering
\caption{FPGA Resource Utilization}
\vspace{-0.05in}
\label{table_2}
{\footnotesize
    \resizebox{\columnwidth}{!}{
    \begin{tabular}{cccccc}
        \toprule
        {\bf  } & {\bf LUT} & {\bf FF} & {\bf BRAM} & {\bf URAM} & {\bf DSP}\\
        \hline\hline
        HLS      &  19362     & 281743    & 818       & 80       & 837\\
         &  (37.37\%) & (26.95\%) & (83.13\%) & (62.5\%) & (42.53\%)\\
        \hline
        RTL      &  259166    & 270267    & 874       & 96       & 783\\
         &  (49.58\%) & (25.85\%) & (88.82\%) & (75.0\%) & (39.79\%)\\
        \bottomrule
    \end{tabular}}
}
\end{table}

\begin{figure*}[t] 
    \centering 
    \includegraphics[scale=0.7]{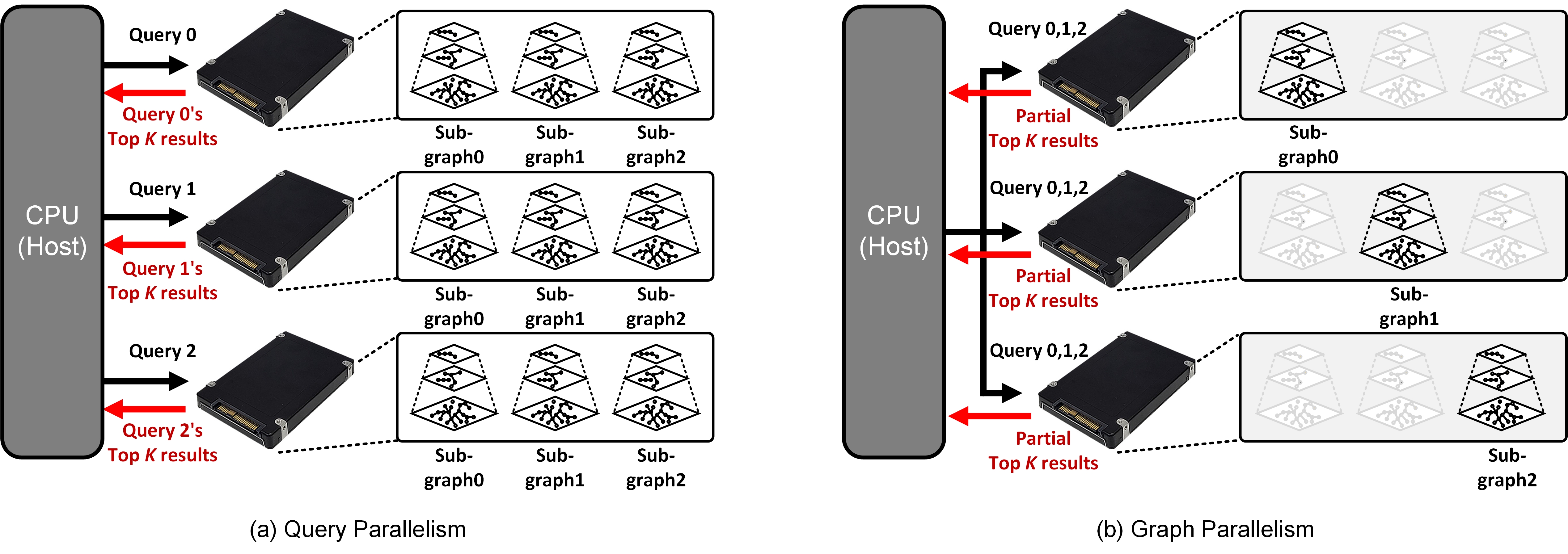} 
    \caption{Two Different Parallelization Methods} 
    \label{fig_10} 
\end{figure*}

\begin{figure*}[t] 
    \centering 
    \includegraphics[scale=0.7]{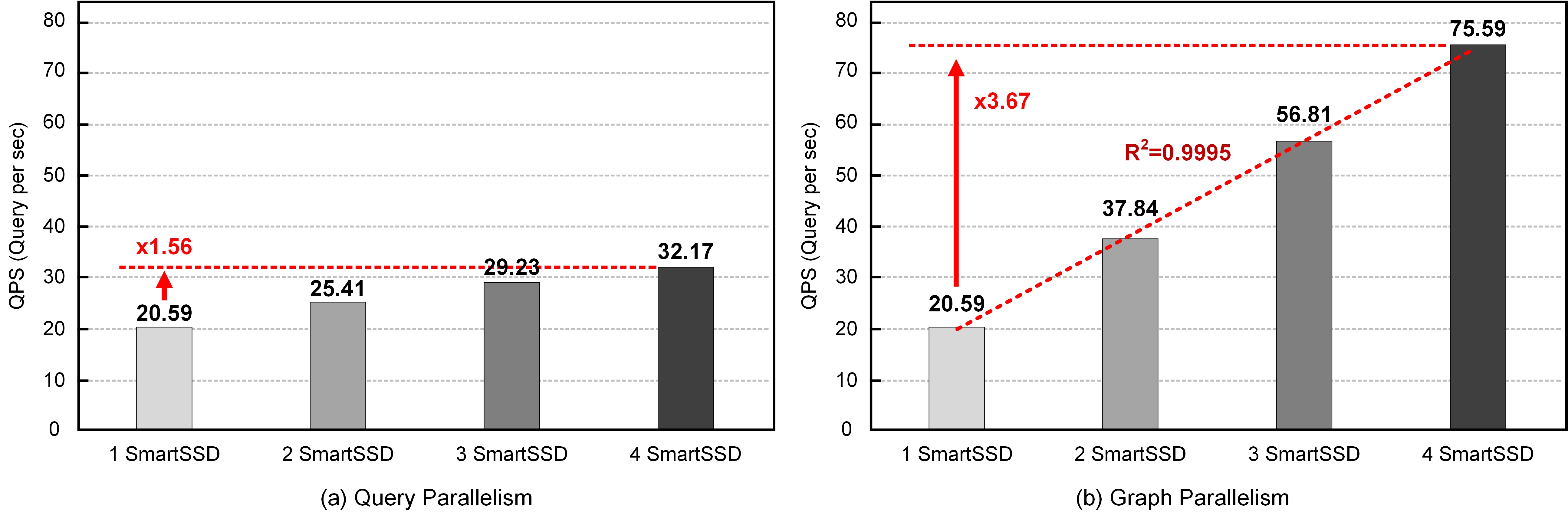} 
    \caption{Scaled-up System Performance of Two Different Parallelization Methods} 
    \vspace{-0.2in}
    \label{fig_11}
\end{figure*}

\begin{figure*}[t] 
    \centering 
    \includegraphics[scale=0.7]{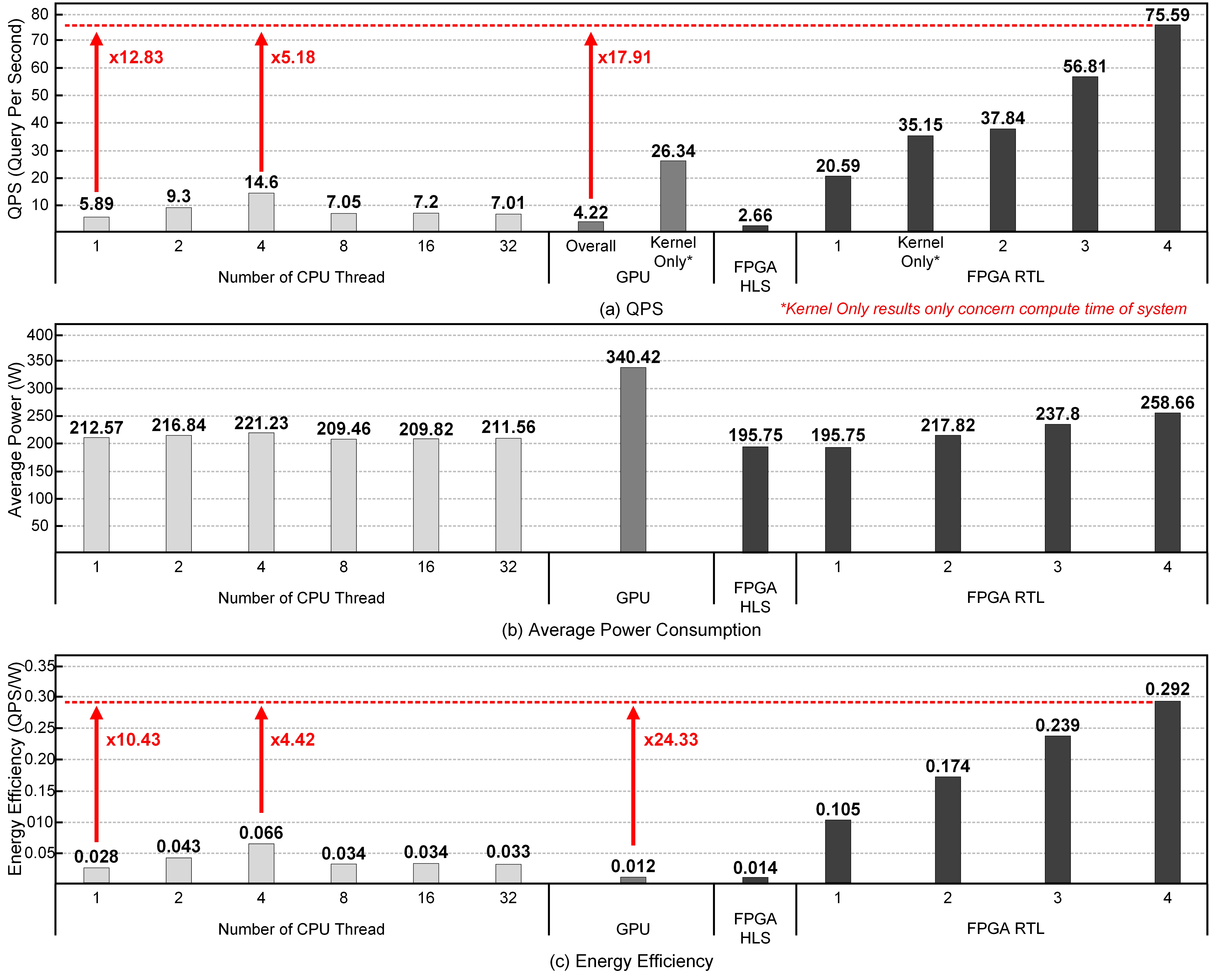} 
    \caption{Large-Scale ANN Performances on Various Server Platforms: (a) QPS (b) Average Power (c) Energy Efficiency} 
    \label{fig_12}
    \vspace{-0.2in}
\end{figure*}

\subsection{Scaling-up with Multiple SmartSSDs}
One major advantage of using SmartSSD is that it is easy to scale out the number of devices with its standard U.2 form factor in the storage server. Managing multiple SmartSSD devices only requires a small revision to the host code. The host needs to identify each device and program it individually. Once programmed, the host creates a device context for each and sends commands through the context as it does in the single device case.
On the modified HNSW algorithm, in which we already divided the large database into smaller graph databases, we have two parallelization strategies to utilize multiple SmartSSDs: query parallelism and graph parallelism.

In query parallelism, each SmartSSD holds a full copy of the graph databases and processes a different set of queries, as shown in Figure \ref{fig_10} (a). In this case, there is no dependency among devices because each SmartSSD has an entire database and works on different queries. However, all sub-graph databases need to be loaded from the SSD to DRAM, so the latency required for database movement can become a bottleneck in the overall system. On the other hand, graph parallelism distributes graph databases among multiple SmartSSDs, without any database duplication. Each device is responsible for finding the top $K$ results out of the assigned graph databases, as shown in Figure \ref{fig_10} (b). Since the number of sub-graphs processed in a single SmartSSD decreases, the database transfer latency is also reduced in proportion to the number of devices. However, the top $K$ results from each SmartSSD need to be aggregated in the host, and the brute-force search is required to compute the final nearest neighbors.

Figure \ref{fig_11} shows the QPS performance of the two different parallelization approaches when the number of SmartSSDs varies from 1 to 4. As shown in Figure \ref{fig_11} (a), the performance of the query parallelism does not scale linearly as the number of devices increases. It achieves 32.17 QPS with 4 SmartSSD devices, which is only 1.56x higher than the single device performance. The reason for the nonlinear increase in performance is mainly because each device still needs to load an entire database from the SSD to DRAM. The computation time does reduce by four times because each divice processes only 1/4 of the input queries, but this means the effect of input batching is also reduced by four times because it has a smaller number of queries to work on with a single memory load. As a result, the database transfer time, which is a bottleneck of the system, hinders the linear increase in the QPS performance.
In comparison, the performance of the graph parallelism increases almost linearly as the number of devices increases. It achieves 75.59 QPS performance, which is 3.67x higher than the single device performance, as shown in Figure \ref{fig_11} (b). Unlike query parallelism, graph parallelism reduces computation time as well as data transfer time by reducing the number of graph databases to cover per device. In graph parallelism, the host should aggregate the results from the devices and perform the last brute-force search. However, the host needs to find just the ten best searches out of 40 candidates for each query. Thus, its workload is negligible. Based on our measurement, it takes 0.29s, which is only 0.2\% of the overall execution time.

\subsection{Evaluation}
To evaluate the performance of a large-scale ANN search on different server platforms, we measure the QPS and average power consumption of each server platform when they run the HNSW algorithm on the SIFT1B dataset with the same configuration ($ef$=40, $K$=10). For the CPU-based server platform, we change the number of threads from 1 to 32 and enable the SIMD extension for efficient vector distance calculations.

Figure \ref{fig_12} shows the QPS, average power consumption, and energy efficiency of the three server platforms. The performance of the CPU-based server platform linearly increases until the number of threads is 4. However, its performance is saturated beyond this point, mainly because of the memory interference among multiple threads. The CPU utilization is also saturated at a low level, implying that the target workload is not compute bound, but memory bound.
We confirm this observation in the GPU setting as well. Although GPU’s compute kernel performance achieves 26.34 QPS with heavy optimizations in CUDA and maximized query batching (i.e., processing 10,000 queries at a time), its overall performance drops to 4.22 QPS. This is because the graph database stored on the system’s SSD should be first copied to the host main memory and then transferred to GPU’s local DRAM for processing.
On the contrary, the proposed computational storage platform solves this data movement problem by placing the high-capacity SSD and the FPGA in the same device and utilizing P2P communication between them for efficient near data processing. The platform achieves 20.59 QPS, which is only 40\% lower than its maximum compute capability of 35.15 QPS.
In addition, the proposed computational storage platform can scale up by harnessing multiple SmartSSD devices and achieve better performance. With 4 SmartSSDs, the platform’s overall performance increases almost linearly to 75.59 QPS.
As a result, the proposed computational storage platform shows 1.41-3.5x and 4.88x higher QPS performance than the CPU-based and GPU-based server platform with a single SmartSSD. More remarkably, with 4 SmartSSDs equipped, the same platform shows 5.18-12.83x, and 17.91x higher QPS performance than the CPU-based and GPU-based server platforms, respectively.

Figure \ref{fig_12} (b) shows the average power consumption of the three server platforms. We confirm that the idle power of the CPU-based, GPU-based server platform and SmartSSD-based computational storage platform without extra computing devices such as GPU and SmartSSD are 200.4W, 194.4W, and 178W respectively.
For the CPU-based server platform, the overall tendency of the power consumption is similar to its QPS results. The power marginally increases up to 221W until the number of threads reaches four but decreases to around 210W after the overall performance is saturated.
The GPU-based server platform and the computational storage platform, each with an extra computing device, consume 340.42W and 195.75W, respectively. Considering the idle power of each platform is 194.4W and 178W, the power-hungry GPU significantly contributes to the high power consumption. On the other hand, the computational storage platform with a single SmartSSD device consumes as much power as that of the CPU-based server running a single or two threads when we equalize the idle power of the two server platforms. 
When the number of SmartSSD devices increases, the power consumption of the computational storage platform, except for the base server power, increases linearly. However, the SmartSSD device itself is much more power-efficient than the GPU, so the computational storage platform with 4 SmartSSDs only consumes 258.66W, which is still significantly less than the GPU-based server.
Finally, the energy efficiency of the three server platforms, which is calculated by QPS performance divided by average power consumption (= QPS/W), is shown in graph (c). The computational storage platform with a single SmartSSD achieves 1.59-3.75x and 8.75x higher energy efficiency than the CPU-based and the GPU-based server platform, respectively. The same platform with 4 SmartSSDs achieves 4.42-10.43x and 24.33x higher energy efficiency than the CPU-based and the GPU-based server platform, respectively. The large performance and energy efficiency gain over the conventional server platforms is achieved because the proposed platform effectively reduces the data transfer overhead at the system level by near-data processing in SmartSSD. The accelerator design for a complex HNSW search algorithm also contributes to the gain.
It is also noteworthy that the SmartSSD’s power consumption and its form factor are about 10x smaller than a comparable GPU card, which makes SmartSSD even more promising for future data center applications.

\subsection{Discussion}
Given the limited resource of hardware and experiment environment, we utilized three different server platforms with fixed configurations to demonstrate the merits of our near-data processing solution. With recent advances in memory (e.g., HBM, NVMe SSD) and technology supported by the hardware (e.g., GPUDirectStorage\cite{NVIDIA}), architects are provided with various sets of hardware and tools for constructing their system. This section discusses some key design parameters of the systems and their effect on overall systems.

\textbf{SSD bandwidth.} While our baseline CPU- and GPU-based server platform uses the SATA interface that provides 0.58GB/s bandwidth, today's NVMe SSDs can support up to 4GB/s of bandwidth via PCIe3x4. Since the storage IO bandwidth is the bottleneck of the overall system, adopting SSDs with higher bandwidth interface makes a baseline server system have comparable throughput to that of SmartSSD. However, the primary advantage of using a computing storage device is the direct integration of the SSD and FPGA within a compact device. This integration allows a large amount of data transfer without going through the long PCIe lanes in the main board of the system. This physical advantage reduces the power consumed in the entire system, and it makes the whole system more energy-efficient.

\textbf{GPU P2P transfer.} GPUDirectStorage (GDS) has recently enabled a direct data path for DMA transfers between GPU memory and storage. This P2P transfer avoids a bounce buffer through the CPU and decreases the latency on data movement. If we use the GDS for P2P communication in our GPU-based server platform, the overall throughput would be mainly determined by the SSD bandwidth. In the current SATA interface, the IO time of the GPU system accounts for 84\% of the total latency. Therefore, when we calculate the expected throughput for the case of NVMe SSD and GDS, the throughput could increase up to 17.28 QPS, which is still lower than the SmartSSD-based server platform. Significant performance improvement can be achieved by larger bandwidth SSD and P2P communication, but GDS still requires the GPU and SSD to be connected via the long PCIe lanes on the mainboard, unlike the SmartSSD based server platform, which definitely causes more power consumption than the proposed platform with SmartSSD.

\textbf{Dataset size.} 5M vectors are the maximum data size that the FPGA can handle at once because we already use 88.82\% of the BRAM in our current implementation. However, using the modified algorithm that divides the monolithic database into multiple databases, our hardware can cover a much larger dataset by iterating through the sub-graphs. For a smaller dataset with less than 5M vectors, a maximum throughput of 4118 QPS can be achieved, which is the throughput when a single graph is executed on the SmartSSD.
\section{Related Works}
There has been rich literature regarding hardware acceleration of nearest neighbor search algorithms. Zhang et al. \cite{zhang2018efficient} presented a novel PQ-based ANN search on the Intel HARPv2 FPGA platform using OpenCL language. Although its proposed PQ-based algorithm is highly parallelizable and scalable with achieving high QPS on FPGA, its low accuracy can be a problem in some cloud services that require high accuracy. Abdelhadi et al. \cite{abdelhadi2019accelerated} implemented a PQ-based ANN search on FPGA to maximize its throughput, but the algorithm innately suffers from low accuracy and targets only a small dataset. There have been many publications that use a computational storage platform for near-data processing. Biscuit \cite{gu2016biscuit} was a platform for near data processing based on a custom board like SmartSSD but with an SSD and low-performance processor. BigStream \cite{samynathan2019computational} targeted big data analytics using the SmartSSD board in PCIe HHHL form factor. However, none of the previous works used a computational storage platform for graph-based ANN search.
\section{Conclusions and Future Works}
In this paper, we present a computational storage platform for fast and energy-efficient billion-scale nearest neighbor search on high-dimensional vector space. Among many approximate nearest neighbor search algorithms, we choose the HNSW algorithm, which is widely used in cloud-scale services due to its high accuracy and database scalability based on the multi-layer graph structure. However, the HNSW algorithm is cumbersome to accelerate in hardware as it involves a huge database and irregular memory access patterns for graph traversing.
In this paper, we first propose to use Samsung’s SmartSSD computational storage device to accelerate the HNSW nearest neighbor search algorithm. The entire database is stored in the SSD while a segment is fetched to a local DRAM for fast access. On FPGA, the HNSW search kernel is implemented in two different design methodologies: HLS and RTL. The HLS-based design heavily modifies the baseline C++ code with various optimizations, while the RTL-based design proposes a custom hardware architecture that can fully utilize the external memory bandwidth with parallelized search logic. 
We also scale up the computational storage platform to have multiple SmartSSD devices. As a result, the proposed platform achieves 75.59 QPS for the SIFT1B dataset, which is up to 12.83x and 17.91x faster, and up to 10.43x and 24.33x more energy efficient than the conventional CPU-based and GPU-based server platform, respectively.
For future works, we plan to improve the performance of the computational storage platform even further by scaling up the system with more SmartSSD devices. We also plan to increase the size of the vector database to tens of TBs so that we can fully utilize the platform’s capability. In the end, we would like to run an end-to-end image search application on the proposed platform. If we can run a data-intensive cloud services on a smaller number of computational storage nodes, it will be very impactful to the cloud industry.


\ifCLASSOPTIONcaptionsoff
  \newpage
\fi

\bibliographystyle{IEEEtranS}
\bibliography{refs}

\end{document}